\documentclass[aps,prb,twocolumn,groupedaddress]{revtex4-1}
\usepackage[bookmarks]{hyperref}
\usepackage[dvips]{graphicx}
\usepackage{amsmath}
\usepackage{amstext}
\usepackage{graphics}
\usepackage{exscale}
\usepackage{epsfig}
\usepackage{changebar}
\usepackage[T1]{fontenc}
\usepackage{bm}
\usepackage{bbm}

\usepackage{version}

\begin{document}
\title[Renormalized perturbation theory for $n$-channel
Anderson model]{Study of Hund's rule coupling in models of magnetic impurities and quantum dots } 
\author{Y Nishikawa${}^{1,2}$  and A C Hewson${}^1$ }
\affiliation{${}^1$Department of Mathematics, Imperial College, London SW7 2AZ,
  United Kingdom.}
\affiliation{${}^2$Graduate School of Science, Osaka City University, Osaka 558-8585, Japan} 

\begin{abstract}
Studies  of the effects of the Hund's rule coupling $J_{\rm H}$ in  multiple orbit impurities or quantum dots using different models have led to  quite different predictions
for the Kondo temperature $T_{\rm K}$ as a function of $J_{\rm H}$.
 We show that the differences depend on whether or not the
 models 
 conserve  orbital  angular momentum about the impurity site.
 Using  numerical renormalization group (NRG)  calculations,  we deduce the
renormalized parameters for the Fermi liquid regime, and show that, despite the
differences between the models,  the low energy fixed point  in the strong
correlation regime is  universal with a single energy scale $T_{\rm K}$,  and just two renormalized interaction
parameters, a renormalized single orbital term, $\tilde U=4T_{\rm K}$, and renormalized
Hund's rule term,  $\tilde J_{\rm H}=8T_{\rm K}/3$.  

\end{abstract}

\maketitle

\section{Introduction}

The role of the Hund's rule coupling $J_{\rm H}$ has been investigated  for magnetic impurities and quantum
dots by a number of authors starting from rather different models \cite{Sch67,NC09,NCH10s,PB05,SNOHT12}. This has led to  different predictions for the
behavior of the Kondo temperature in these systems as a function of $J_{\rm H}$. As a result it is not clear whether or
not  the models may also differ in their predictions for their low energy behavior.
We investigate this question by applying a combination of the numerical renormalization group (NRG) and renormalized perturbation theory (RPT) to some of these different models. From an analysis of single particle excitations about the 
NRG low energy fixed point we can obtain an accurate estimate of the Kondo temperature in all cases.
We can also calculate the  renormalized interaction parameters at this fixed point and compare them for the different models. This information can be used to deduce the low temperature specific heat coefficient, spin and charge susceptibilities
and low energy dynamical behavior for the models in all parameter regimes.\par

A model of a magnetic impurity with $n$-fold degenerate orbitals interaction with a bath of
conduction electrons will in general require quite a large number of parameters
to describe the interactions between the electrons in these orbitals for various filling factors,
and their interaction with the conduction electrons of the host metal.
 To understand the basic physics of these models a number of different models,
using a restricted set of parameters, have been used. 
One of these models was introduced
by Yoshimori (model 1) where the interactions between electrons in the impurity orbitals were specified by
just two parameters \cite{Yos76}, a local Coulomb interaction $U$ and an inter-orbital exchange interaction   $J_{\rm H}$. 
The Hamiltonian of the model takes the form,
\begin{equation} {\cal H}_{\rm\, model\, 1}={\cal H}_0+{\cal H}_d\end{equation}
  \begin{eqnarray}
{\cal H}_0=&\sum_{m\sigma}\epsilon_{dm\sigma}d^{\dagger}_{m\sigma}d^{}_{m\sigma}+\sum_{k,m\sigma}\epsilon_{km\sigma}
c^{\dagger}_{k m\sigma} c^{}_{k m\sigma}\nonumber \\
&+\sum_{k m\sigma} (V_k d^{\dagger}_{m\sigma} c^{}_{k m\sigma}
+  V_k^* c^{\dagger}_{k m\sigma} d^{}_{ m\sigma})
\label{model1a}
\end{eqnarray}
where $d^{\dagger}_{ m\sigma}$, $d^{}_{ m\sigma}$, are creation and
annihilation operators for an electron in an impurity state  with total angular momentum
quantum number 
$l$, and $z$-component $m=-l,-l+1,\cdots, l$, and spin
component
$\sigma=\uparrow,\downarrow$. The impurity level
 in a magnetic field $H$ we take as $\epsilon_{dm\sigma}=\epsilon_d-\mu_{\rm
   B}\sigma H-\mu_{\rm B} mH-\mu$, where  $\sigma=1$ ($\uparrow$) and  $\sigma=-1$ ($\downarrow$) and $\mu$
 is the chemical potential, and $\mu_{\rm B}$ the Bohr magneton.  
The creation and annihilation operators $c^{\dagger}_{km\sigma}$, $c^{}_{km\sigma}$ are
for  partial wave conduction electrons with energy
$\epsilon_{km\sigma}$. The  hybridization matrix element for impurity levels with the
conduction electron states is $V_k$. We denote the hybridization width
factor by $\Delta_{m\sigma}(\epsilon)=\pi\sum_k|V_k|^2\delta(\epsilon-\epsilon_{km\sigma})$,
which we can take to be a constant $\Delta$ in the wide flat band limit.
The remaining part of the Hamiltonian,  ${\cal H}_d$ describes the interaction between
the electrons in the impurity state, 
\begin{eqnarray}
{\cal H}_d=&&
    {(U-J_{\rm H})\over 2}\sum_{mm'\sigma\sigma'} d^{\dagger}_{m\sigma}
    d^{\dagger}_{m'\sigma'} d^{}_{ m'\sigma'} d^{}_{ m\sigma}\nonumber  \\
&&+{J_{\rm H}\over 2}\sum_{mm'\sigma \sigma'} d^{\dagger}_{ m\sigma}
d^{\dagger}_{m'\sigma'} d^{}_{m\sigma'} d^{}_{ m'\sigma}. 
\label{model1b}
\end{eqnarray}
 This model can be used to describe
transition metal impurities, such as Fe or Mn,  in a metallic host in the
absence of spin orbit or crystal field splittings. 
The model can be interpreted more generally 
with $\alpha=m+l+1$ as a channel index taking values $\alpha=1,2, \cdots, n$ where
$n=2l+1$ is the number of channels. The Hund's rule term tends to align the
electrons on the impurity site such that for large $U$ and large $J_{\rm H}$
 the impurity state will correspond to a spin $S=n/2$.
  \par

A different $n$-fold model has been studied recently by Nevidomskyy and Coleman \cite{NC09} (model 2) which also includes
a Hund's rule interaction. This model is characterized by just two interaction terms,
an inter-orbital Hund's rule exchange interaction $ J_{\rm H}$ 
and an Kondo form of exchange interaction  $ J_{\rm K}$ between the impurity electrons and the
conduction electrons, so that the Hamiltonian takes the form,
\begin{eqnarray}H_{\rm\, model\,2}=&\sum_{k,\alpha\sigma}\epsilon_{k\alpha\sigma}
c^{\dagger}_{k \alpha\sigma} c^{}_{k \alpha\sigma}-J_{\rm H}\large(\sum_{\alpha=1}^n{\bf
  s_\alpha}\large)^2 \nonumber \\
&+J_{\rm K}\sum_{\alpha=1}^n\sum_{k,k'}{\bf S_\alpha}.c^{\dagger}_{k
  \alpha\sigma}{\bf \sigma}_{\sigma,\sigma'} c^{}_{k' \alpha\sigma'}
\label{model2}\end{eqnarray}
There are significant differences between  models 1 and 2, which can be seen
more clearly if we restrict the discussion to the case $n=2$, when model 1
 can be written in a more usual form with the exchange interaction
written in terms of spin operators,
\begin{equation}
{\cal H}_{d,\rm \,model\,3}=U\sum_{\alpha=1,2}n_{\alpha\uparrow}n_{\alpha\downarrow}
   +U_{12}\sum_{\sigma\sigma'}n_{1\sigma}n_{2\sigma'}
-{2J_{\rm H}} {\bf s}_{1}\cdot{\bf s}_{2}.  
\label{model3}
\end{equation}
with  $U_{12}=U-3J_{\rm H}/2$. We generalize the model by taking  $U_{12}$
as an independent parameter, and refer to the model in this generalized form
as  model 3. It has been used in this form to  describe the interactions in
certain double quantum dots\cite{SNOHT12}.\par
We consider,   first of all, model 3 in the limit $J_{\rm H}=0$ with $U_{12}=0$.
In this case there is no interaction or hybridization between the two
channels and the model is equivalent to two independent Anderson models.
In the strong coupling regime $U/\pi\Delta\gg 1$
 with particle-hole symmetry it can be converted
into model 2 ($J_{\rm H}=0$)  via a Schrieffer-Wolff
transformation with $J_{\rm K}=4V^2/U$.  If the first order correction
term in $J_{\rm H}$ is included in this limit, then model 3
can be seen to be essentially equivalent to model 2 in Eq. (\ref{model2})
for $n=2$.\par

However, for  $J_{\rm H}=0$, and $U_{12}=U$ in Eq. (\ref{model3}), corresponding
to the Yoshimori model (model 1), there is an interaction between the two
channels. In this case the orbital index $l$ can be combined with the spin
index into a single index $(m,\sigma)$ running over 4 values and the
 model can be shown to have  SU(4) symmetry, corresponding to conservation of both
spin and orbital angular momentum. In the particle-hole localized limit
with $U/\pi\Delta\gg 1$, it can be transformed via a Schrieffer-Wolff
transformation (which takes into account virtual excitations from the 2-electron,
six-fold degenerate,
ground state of the impurity to  1-electron and 3-electron excited
states) into an SU(4) Kondo model with $J_{\rm K}=4V^2/U$
with a six dimensional representation for the local SU(4)
operators\cite{NCH10s}. Hence models 1 and 2 are quite different in the
Kondo regime for $J_{\rm H}=0$. For general $n$, in the Kondo regime when  $J_{\rm H}=0$, model 1 reduces
to an SU(2n) Kondo model and model 2 to $n$ independent SU(2) Kondo models.
This reflects the important difference that model 1 is invariant under  orbital
as well as spin rotation, whereas model 2, and model 3
when $U_{12}\ne U-3J_{\rm H}/2$, are invariant under spin rotation only.
This can be verified explicitly by considering the commutator of the generator,
$L^+=\sum_\sigma
(d^{\dagger}_{1,\sigma}d^{}_{2,\sigma}+c^{\dagger}_{k,1,\sigma}c^{}_{k,2,\sigma})$,
with the Hamiltonian of model 3.
This difference could have consequences for the low energy behavior of these
models which we consider in the next section. \par

\section{Low energy Fermi liquid regime}
The conservation of both angular momentum and spin in model 1 permitted 
Yoshimori \cite{Yos76} to derive two Ward identities for interaction vertex part
at zero frequency.  This interaction vertex can be interpretated in terms
of a renormalized interaction between the quasiparticles of a Fermi liquid\cite{NCH10s}
described by the Hamiltonian,
 $\tilde {\cal
  H}=\tilde{\cal H}_0+\tilde{\cal H}_{d}$, where
\begin{eqnarray}
&&\tilde{\cal H}_0=\sum_{m,\sigma}\tilde\epsilon_{d,m} \tilde d^{\dagger}_{m,\sigma}\tilde d^{}_{m,\sigma}+\sum_{k\sigma}\epsilon_{k,m}
c^{\dagger}_{k,m,\sigma} c^{}_{k,m,\sigma}\nonumber \\
&&+\sum_{k,m,\sigma} (\tilde V_k \tilde d^{\dagger}_{m,\sigma} c^{}_{k,m,\sigma}
+ \tilde V_k^* c^{\dagger}_{k,m,\sigma} \tilde d^{}_{ m,\sigma})
 \label{qpmodel1a}
\end{eqnarray}
and

\begin{eqnarray}
\tilde {\cal H}_d=&&
    {(\tilde U-\tilde J_{\rm H})\over 2}\sum_{mm'\sigma\sigma'} :\tilde d^{\dagger}_{m\sigma}
   \tilde d^{\dagger}_{m'\sigma'}\tilde d^{}_{ m'\sigma'}\tilde d^{}_{ m\sigma}:\nonumber  \\
&&+{\tilde J_{\rm H}\over 2}\sum_{mm'\sigma \sigma'}:\tilde d^{\dagger}_{ m\sigma}
\tilde d^{\dagger}_{m'\sigma'}\tilde d^{}_{m\sigma'} \tilde d^{}_{ m'\sigma}:. 
\label{qpmodel1b}
 \end{eqnarray}
This effective model is of the same form as the original model defined in Eqs.
(\ref{model1a}) and  (\ref{model1b}) with the difference that
the interaction terms have to be normal ordered. The brackets $:{\hat O}:$ indicate the normal
ordering of the operator
${\hat O}$  with respect to the ground state of
the interacting system, which  plays the role of  the vacuum. This term
only comes into play when more than one quasiparticle is created from the vacuum. \par
From the  Ward identities\cite{Yos76,YZ82,
NCH10s}
 exact expressions can be derived  for the impurity
charge, spin and orbital susceptibilities, $\chi_c$, $\chi_s$,
and $\chi_{orb}$, at $T=0$,
\begin{equation}
\chi_c=2n\tilde \rho^{(0)}(0)(1-((2n-1)\tilde U-3(n-1)\tilde J_{\rm H})\tilde\rho^{(0)}(0)).
\label{chic}
\end{equation}
\begin{equation}
\chi_s=2n\mu_{\rm B}^2\tilde \rho^{(0)}(0)(1+(\tilde U+(n-1)\tilde J_{\rm H})\tilde\rho^{(0)}(0)),
\label{chis}
\end{equation}
\begin{equation}
\chi_{orb}={(n^2-1)\mu_{\rm B}^2\tilde \rho^{(0)}(0)\over 12}\left(1+(\tilde U-3\tilde J_{\rm H})\tilde\rho^{(0)}(0)\right),
\label{chiorb}
\end{equation}
where $\tilde\rho^{(0)}(\omega)$ is the free 
quasiparticle density of states per single spin and channel,
\begin{equation}
\tilde \rho^{(0)}(\omega)={\tilde\Delta/\pi\over (\omega-\tilde\epsilon_{d,\alpha})^2 +\tilde\Delta^2}.\label{fqpdos}
\end{equation} 
The impurity contribution to the  specific heat coefficient $\gamma$ is also
 given exactly  by
 $\gamma=2n\pi^2\tilde \rho^{(0)}(0)/3$ and the Wilson ratio $R_{\rm W}=
\pi^2\chi_s/3\mu^2_{\rm B}\gamma$  is given by 
\begin{equation}
R_{\rm W}=1+(\tilde U+\tilde J_{\rm H})\tilde \rho^{(0)}(0).\label{WR}
\end{equation}
\par
In  the localized
 limit $U/\pi\Delta\gg 1$ with particle-hole symmetry ($\tilde\epsilon_d=0$), the charge susceptibility is suppressed
so we can equate  $\chi_c$ to zero. Similarly when $J_{\rm H}$ is large  the orbital
susceptibility will  be suppressed so that $\chi_{orb}$ can also be equated to
 zero. 
 Then from  Eqs. (\ref{chic}) and  (\ref{chiorb}) 
 we obtain for
 the spin susceptibility,
\begin{equation}
\chi_s={(g\mu_{\rm B})^2S(S+1)\over 3T_{\rm K}},
\label{chis2}
\end{equation}
where $S=n/2$ and $g=2$, and the relation between the renormalized parameters,
\begin{equation}
\pi\tilde\Delta=\tilde U={3\over 2}\tilde J_{\rm H}=4T_{\rm K}.\label{relation2}
\end{equation}

As angular momentum is not conserved for model 2, and model 3 if
$U_{12}\ne U-3J_{\rm H}/2$, the question arises as to
whether the low energy fixed point of this model can be described by a
quasiparticle Hamiltonian  with renormalized parameters, similar to that given in Eqs. (\ref{qpmodel1a}) and
(\ref{qpmodel1b}), and if so,  
does the relation given in Eq. (\ref{relation2}) hold when $J_{\rm
  H}$ is large? \par

To examine this question we  extend our earlier NRG calculations\cite{NCH10s}  of the renormalized parameters
for model 1 to models 2 and 3 for the case $n=2$.
 We can test the hypothesis that the low temperature
behaviour of all three models can be described by a quasiparticle
Hamiltonian in terms a set of renormalized parameters, $\tilde\Delta$, $\tilde
U$, $\tilde U_{12}$ and $\tilde J_{\rm H}$.
We restrict our calculations to the
particle-hole
symmetric case and take $\epsilon_d=-U/2-U_{12}$ in the one-electron part of the 
Hamiltonian given in Eq. (\ref{model1a}). In the NRG calculations we used for the
 discretization parameter $\Lambda=6$ and half-bandwidth $D=1$, and approximately 4000
states were retained at each iteration. The value $\pi\Delta=0.01$ was 
used for models 1 and 3.
\par

\section{Calculation of renormalized parameters and $T=0$ Susceptibilities}

The renormalized parameters that describe the quasiparticles and their
interactions can be deduced from an analysis of the low energy
NRG fixed point. The parameters $\tilde\epsilon_d$ and $\tilde\Delta$
can be deduced by fitting the lowest single  particle and hole excitations
from the 
NRG ground state to those of a non-interacting Anderson model.  The interaction parameters, $\tilde U$, $\tilde U_{12}$
and $\tilde J_{\rm H}$, can be deduced from the lowest two-particle excitations, in the same
channel for  $\tilde U$, and  different channels for   $\tilde U_{12}$
and $\tilde J_{\rm H}$. We have analysed the fixed point in the same way for 
all the models, including model 2. Details of these calculations were given in earlier
work\cite{HOM04,NCH10s, NCH10a,NCH12b} so we
shall just quote the results of such an analysis here.\par
 
\begin{figure}[!htbp]
   \begin{center}
     \includegraphics[width=0.47\textwidth]{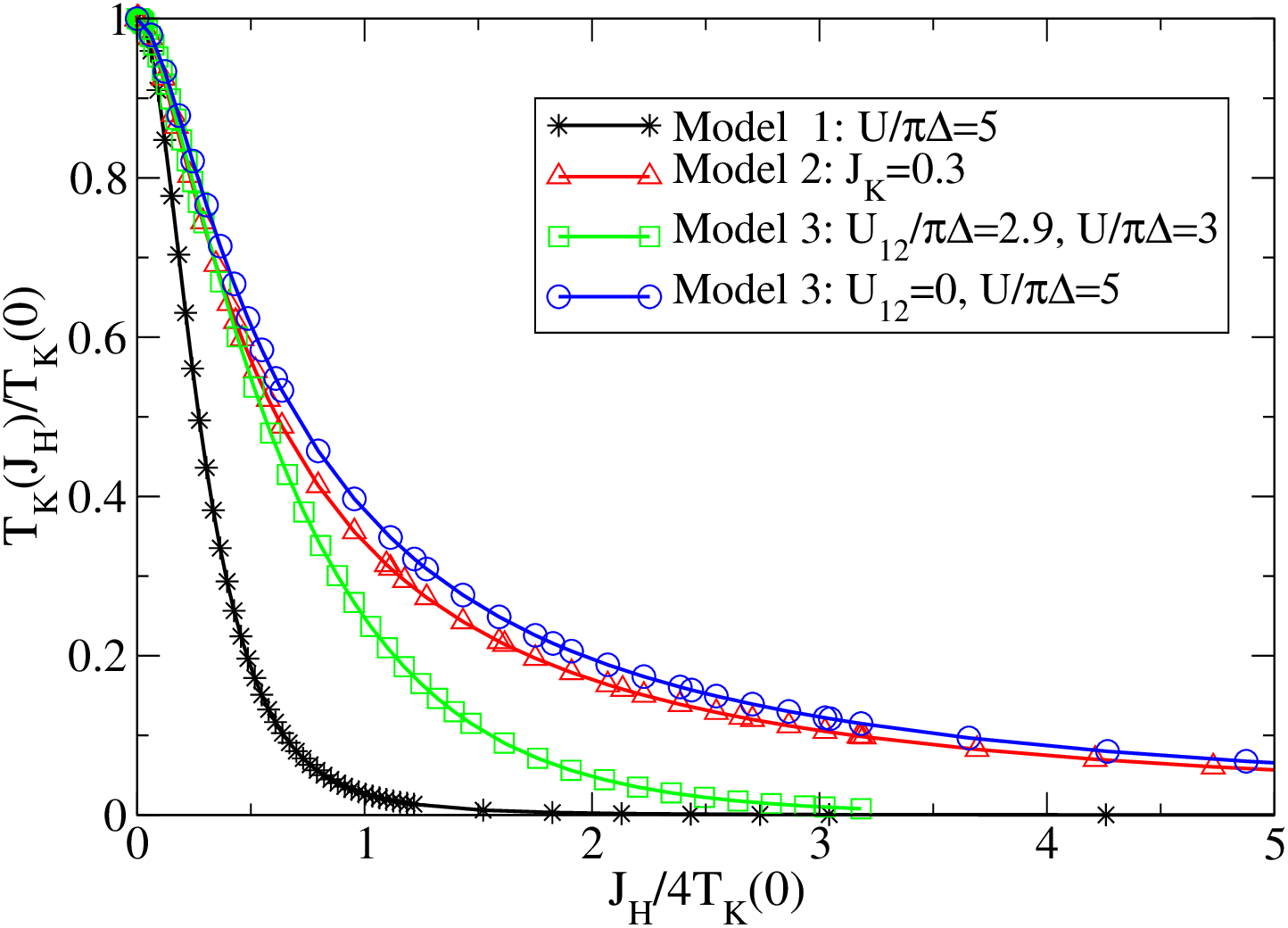}
     \caption{(Color online) A plot of $T_{\rm K}(J_{\rm H})/T_{\rm K}(0)$ versus $J_{\rm
         H}/4T_{\rm K}(0)$  for model 1 with $U/\pi\Delta=5$, model 2 with $J_{\rm K}=0.3$, model 3 with $U_{12}/\pi\Delta=2.9$, $U/\pi\Delta=3$ and also with   $U_{12}/\pi\Delta=0$, $U/\pi\Delta=5$. } 
     \label{rdel}
   \end{center}
 \end{figure}

We use the relation, $\pi\tilde\Delta=4T_{\rm K}$ to deduce the Kondo
temperature in the particle-hole symmetric case for all the models.
The results for $T_{\rm K}(J_{\rm H})/T_{\rm K}(0)$ as a function of $J_{\rm
  H}/4T_{\rm K}$ are shown in Fig. \ref{rdel} for particular parameter
sets for the three models. For model 1 with $U/\pi\Delta=5$, the value of $T_{\rm K}(J_{\rm
  H})$ falls of rapidly with increase of $J_{\rm H}$. Results for this case
were presented in earlier work\cite{NCH10s}, and shown to decrease
exponentially with a fit $T_{\rm K}/\pi\Delta=0.0854{\rm exp}(-1.49\pi^2J_{\rm
  H}/\pi\Delta)$ with $\pi\Delta=0.01$. This exponential dependence is to be
expected  from a Schrieffer-Wolff transformation in the regime
when $U/\pi\Delta\gg 1$ and  $J_{\rm H}$ is large, when the ground state of the
isolated impurity is a two electron spin triplet state. When virtual
excitations to local states with one more or one less electron are taken into
this leads to a spin 1 Kondo model with an effective antiferromagnetic
coupling
$J^*_{\rm K}=4V^2/(U+3J_{\rm H})$,
 giving exponential factor in $J_{\rm H}$ in the resulting
Kondo temperature $T_{\rm K}^*\sim De^{-1/2J^*_{\rm K}\rho_c}$,
where $\rho_c$ is the density of states of the conduction electrons at the
 Fermi level. \par
 The results for model 2 are for the case with
$J_{\rm K}=0.3$, and are seen to decrease much more slowly with increase
of $J_{\rm H}$ and appear to level off for larger values of $J_{\rm H}$.
 However, the initial values of $T_{\rm K}(0)$ are rather different for the
 two models. As model 2 in this limit reduces to $n$-independent SU(2) Kondo models
 the Kondo temperature is given by $T_{\rm K}(0)\sim De^{-1/2J_{\rm K}\rho_c}$. In this limit, on the other hand, model 1 corresponds
to an SU(2n) Kondo model, which  has a Kondo temperature $T_{\rm K}\sim D
e^{-1/2nJ_{\rm K}\rho_c}$, so for the same coupling $T_{\rm K}(0)$ is likely to be
significantly larger than that for the SU(2) model.  
\par
The results for model 3
with $U/\pi\Delta=5$, $U_{12}=0$, also shown in Fig. \ref{rdel},
are very similar to those of model 2. This is not surprising because
because, as noted earlier, when $U_{12}=0$ and $J_{\rm H}=0$ it can be transformed into model
2 (with $J_{\rm H}=0$) via a  Schrieffer-Wolff transformation when
$U/\pi\Delta$ is large, and
 adding the first order correction term
in $J_{\rm H}$ gives essentially model 2.\par
Another parameter set for model 3 with $U/\pi\Delta=3$, $U_{12}/\pi\Delta=2.9$,
 is shown in Fig. \ref{rdel}. In this case there is still a marked fall
off of $T_{\rm K}$ with  $J_{\rm H}$ but intermediate between
the other cases shown.\par

\vspace*{0.7cm}
 \noindent

\begin{figure}[!htbp]
   \begin{center}
     \includegraphics[width=0.47\textwidth]{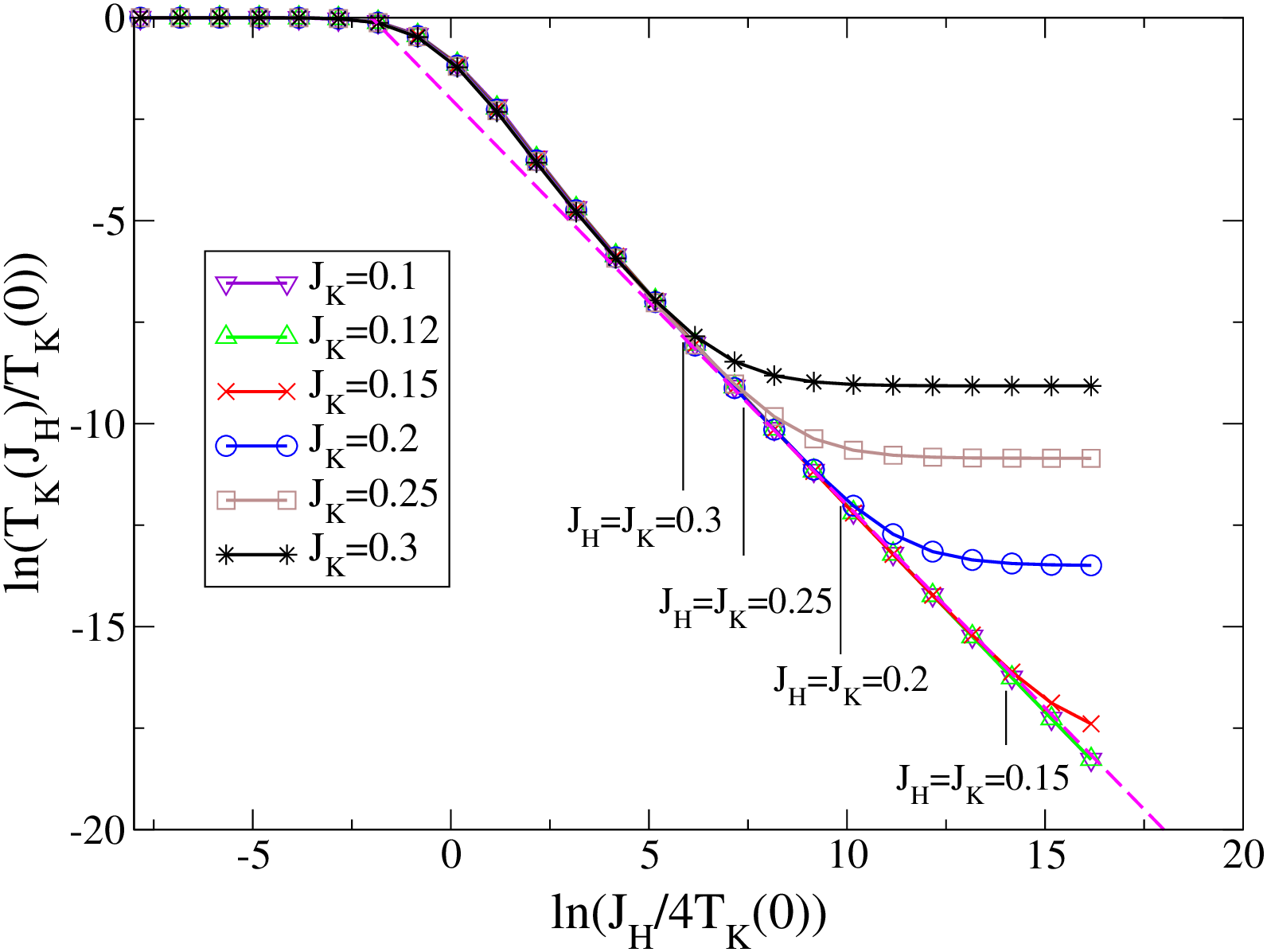}
     \caption{(Color online) The graph shows a plot of ${\rm
         ln}(T_{\rm K}(J_{\rm H})/T_{\rm K}(0))$
       versus  ${\rm ln}(J_{\rm H}/4T_{\rm K}(0))$ for model
       2 for values of $J_{\rm K}=0.1,0.12,0.15,0.2, 0.25,0.3$. 
The dashed line corresponds to $T_{\rm K}(J_{\rm H})={4T^2_{\rm K}(0)/
  e^2J_{\rm H}}$, on using the relation  $4T_{\rm K}(J_{\rm
  H})=\pi\tilde\Delta(J_{\rm H})$. The deviation from the approximate linear
  dependence to a plateau region for $J_{\rm K}=0.15,0.2,0.25,0.3$  occurs when $J_{\rm H}\approx
  J_{\rm K}$ in all four cases.}
     \label{log_rpidel}
   \end{center}
 \end{figure}

In Fig. \ref{log_rpidel} we examine the dependence of $T_{\rm K}$ on $J_{\rm H}$
         for model 2 in more detail by  plotting  ${\rm
         ln}(T_{\rm K}(J_{\rm H})/T_{\rm K}(0))$
       versus  ${\rm ln}(J_{\rm H}/4T_{\rm K}(0))$ 
  for a range of values for the Kondo coupling, $J_{\rm
         K}=0.1,0.12,0.15,0.2, 0.25,0.3$. All the results fall on a
         universal
curve provided $J_{\rm K}<J_{\rm H}$. At $J_{\rm K}\approx J_{\rm H}$
the curves begin to deviate and develop a plateau for  $J_{\rm
         H}>J_{\rm K}$ for the larger values of   $J_{\rm
         K}$, $J_{\rm
         K}=0.15, 0.2, 0.25,0.3$. Plateaus also develop for 
the smaller values of  $J_{\rm
         K}$  for $J_{\rm
         H}/T_{\rm K}(0)$ beyond the range shown in Fig. \ref{log_rpidel}.  We can compare these
         results with the scaling relation derived to one loop order
for this model  by 
Nevidomskyy and Coleman \cite{NC09} (Eq. (11) in their paper with $n=2$)
for intermediate values of the Hund's rule coupling $J_{\rm H}$, which predicts $T_{\rm K}(J_{\rm H})
\propto T^2_{\rm K}(0)/J_{\rm H}$.  This would correspond to a linear behavior
in the log-log plot in Fig. \ref{log_rpidel} with a slope $-1$. The dashed
         line  Fig. \ref{log_rpidel} corresponding to $T_{\rm K}(J_{\rm H})={4T^2_{\rm K}(0)/
  e^2J_{\rm H}}$ does fit the curves over a substantial range, particularly
for the smaller values of $J_{\rm K}$. However, a linear fit to the range $J_{\rm
         H}<J_{\rm K}$ for larger values of $J_{\rm K}$ would give a larger
         slope
with  $T_{\rm K}(J_{\rm H})
\propto (J_{\rm H})^{-x}$, with $x$ varying $1<x<1.25$ depending on the fitting
         range chosen.  The linear fit with $x\approx 1$ 
         begins  at values of $J_{\rm 
         H}\sim 4T_{\rm K}(0)$.\par

\vspace*{0.7cm}
 \noindent

\begin{figure}[!htbp]
   \begin{center}
     \includegraphics[width=0.47\textwidth]{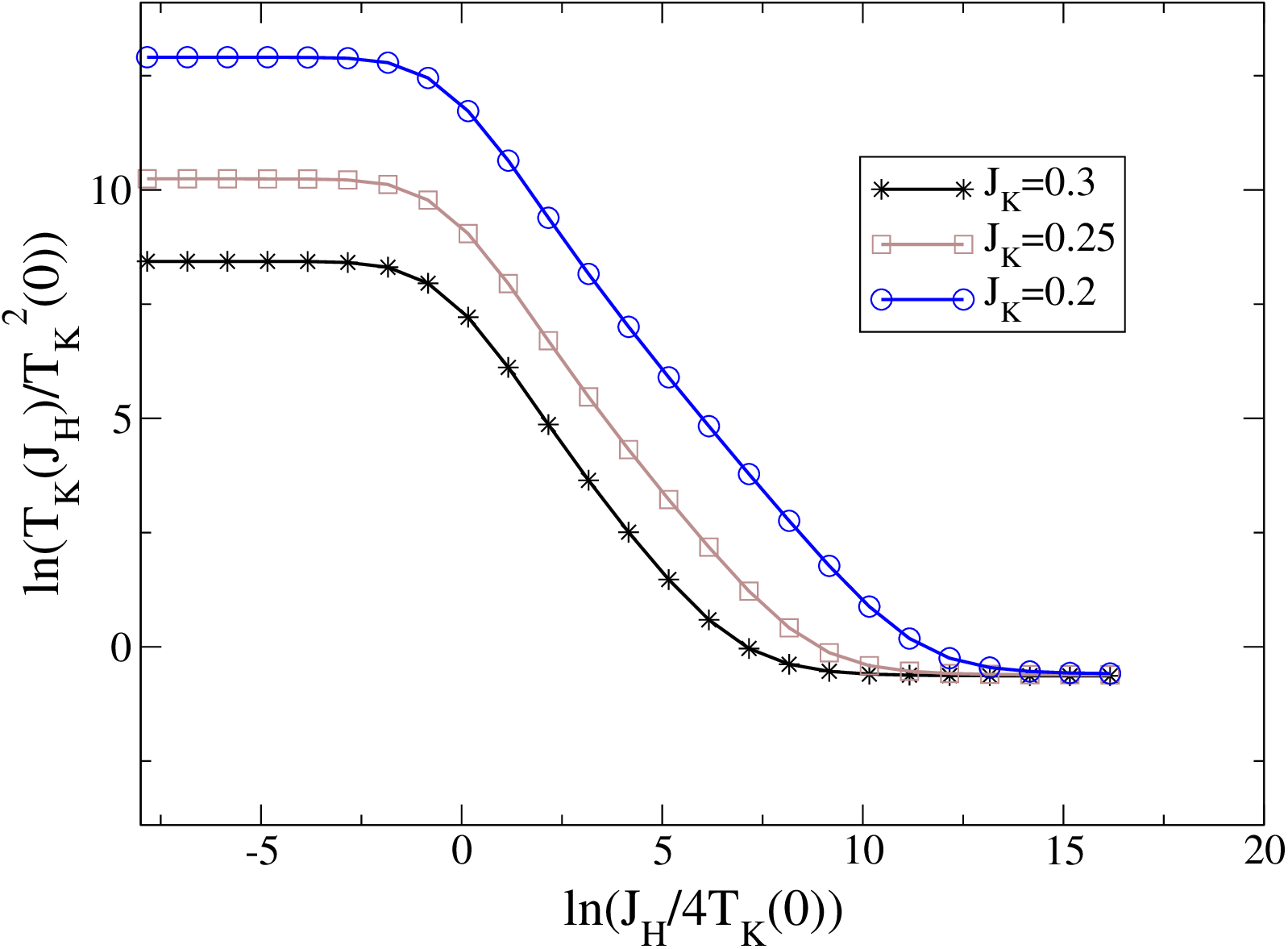}
     \caption{(Color online) A plot of ${\rm
         ln}(T_{\rm K}(J_{\rm H})/T_{\rm K}^2(0))$
       versus  ${\rm ln}(J_{\rm H}/4T_{\rm K}(0))$ for model
       2 for values of $J_{\rm K}=0.3,0.25,0.2$. The results indicate that
       the ratio $T_{\rm K}(J_{\rm H})/T_{\rm K}^2(0)$ becomes independent of $J_{\rm K}$
in the regime  $J_{\rm H} > J_{\rm K}$.
}
     \label{2log_rpidel}
   \end{center}
 \end{figure}

When $J_{\rm H}\gg J_{\rm K}$ and the two impurity spins 
are locked together, model 2 should also be equivalent to an effective Kondo model with the electrons in the  two channels coupled to a spin
$S=1$.  For the general $n$-channel model this Hamiltonian would take the form,
\begin{equation}
H= J^*_{\rm K}\sum_{m=1}^n\sum_{k,k'}{\bf S}.c^{\dagger}_{k m\sigma}{\bf \sigma}_{\sigma,\sigma'} c^{}_{k' m\sigma'}\end{equation}
with an effective coupling  $J^*_{\rm K}$ of a spin $S=n/2$ to the
conduction electrons. 
In the limit $J_{\rm H}\to \infty$ the result, $J^*_{\rm
  K}=J_{\rm K}/n$, derived  by Schrieffer \cite{Sch67} for model 2 should be valid. The
Kondo temperature 
of the model with $n=2S$ would then be given by $T_{\rm K}^*\sim De^{-1/2J^*_{\rm K}\rho_c}$,
neglecting any dependence of the prefactors on $J^*_{\rm K}$.
 The
corresponding result for the model with $J_{\rm H}=0$, $T_{\rm
  K}(0)\sim De^{-1/2J_{\rm K}}\rho_c$, again neglecting any dependence of the
prefactors on $J_{\rm K}$,  would then
imply $T_{\rm K}^*\propto  (T_{\rm K}(0))^n$.  We can test this result for
$n=2$  by modifying the plot given in Fig. \ref{log_rpidel} and plot 
instead  ${\rm
         ln}(T_{\rm K}(J_{\rm H})/T_{\rm K}^2(0))$
  versus  ${\rm ln}(J_{\rm H}/4T_{\rm K}(0))$.
The results are shown in Fig. \ref{2log_rpidel}  for $J_{\rm
       K}=0.3,0.25,0.2$, 
It can be seen for the cases shown   that, in the regime  $J_{\rm H}>
J_{\rm K}$, the ratios of $T_{\rm K}(J_{\rm H})/T_{\rm K}^2(0)$ 
 become independent of the value of $J_{\rm K}$ which implies that, for  $J_{\rm H}\gg
J_{\rm K}$,  $T_{\rm K}^*\propto  (T_{\rm K}(0))^2$. This gives a clear criterion
$J_{\rm H}>
J_{\rm K}$ for the Schrieffer relation\cite{Sch67}, $J^*_{\rm
  K}=J_{\rm K}/n$,   to hold.\par

\vspace*{0.7cm}
 \noindent

\begin{figure}[!htbp]
   \begin{center}
     \includegraphics[width=0.47\textwidth]{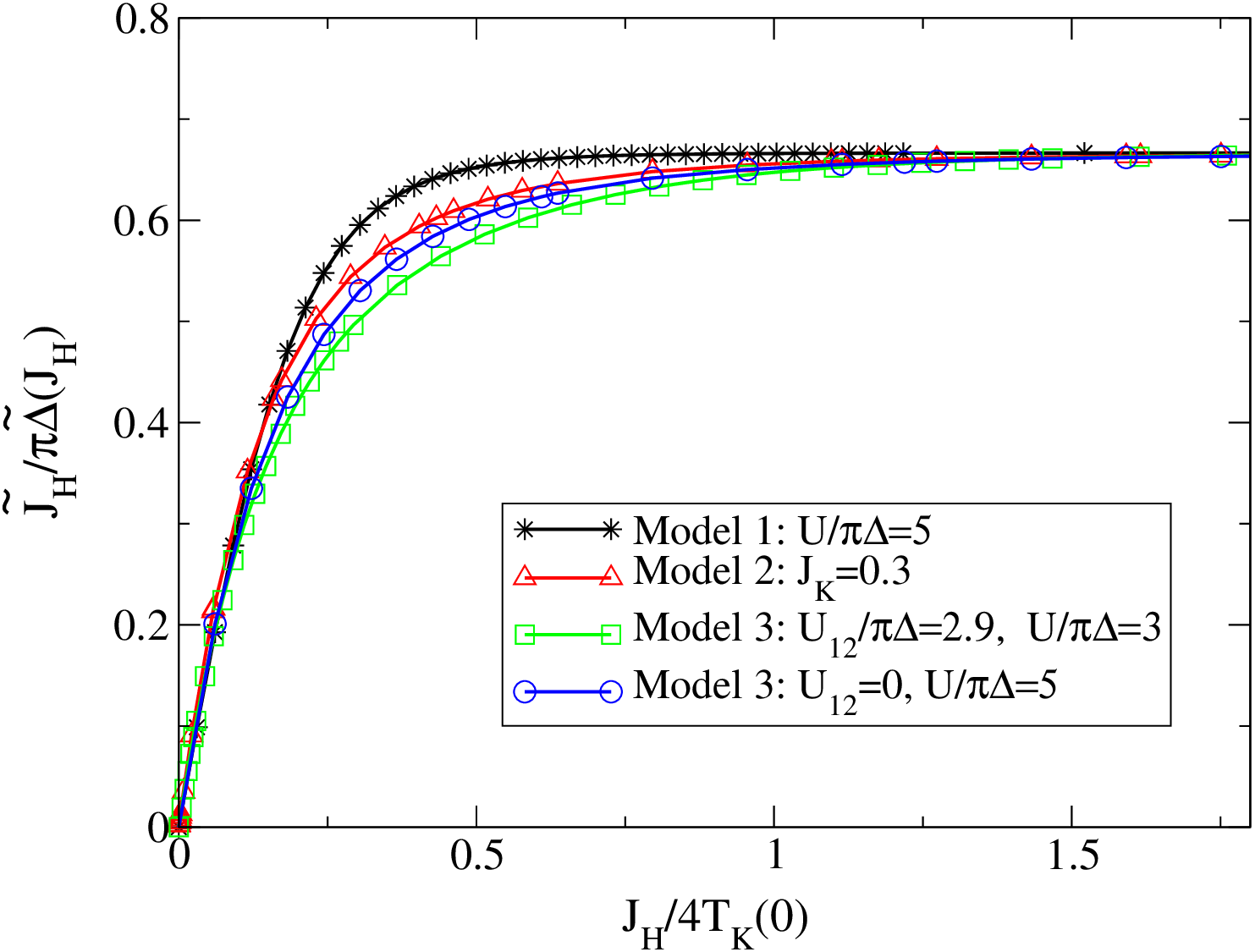}
     \caption{(Color online) A plot of $\tilde J_{\rm H}/\pi\tilde\Delta(J_{\rm H})$
versus   $J_{\rm H}/4T_{\rm K}(0)$ for the models and parameter sets given in Fig. \ref{rdel} } 
     \label{rj}
   \end{center}
 \end{figure}

In Fig. \ref{rj} we plot the ratio of the renormalized parameters, $\tilde
J_{H}(J_{\rm H})/\pi\tilde\Delta(J_{\rm H})$ versus  $
J_{H}/4T_{\rm k}(0)$ for the different models for the  parameter sets given in
Fig. \ref{rdel}. We can see that  all the curves  asyptotically approach the value
predicted in Eq. (\ref{relation2}), $\tilde
J_{H}(J_{\rm H})/\pi\tilde\Delta(J_{\rm H})\to 2/3$ on increasing $J_{\rm H}$. Though
they approach this value at different rates they are all very close to the
limiting value for $J_{\rm H}>5T_{\rm K}(0)$. This clearly shows that the
crossover to $T=0$ susceptibility corresponding to a spin 1 Kondo model,
as given in Eq. (\ref{chis2}),
on increasing $J_{\rm H}$ occurs on a scale $T_{\rm K}(0)$. The regime  $
5T_{\rm K}< J_{\rm H}
< J_{\rm K}$ for model 2 corresponds to the scaling regime  $T_{\rm K}(J_{\rm H})\propto
1/J_{\rm H}$ of Nevidomskyy and Coleman \cite{NC09} where the low temperature behaviour
can be described by a spin 1 Kondo model before the crossover to the regime
$ J_{\rm H}>J_{\rm K}$
where   $T_{\rm K}(J_{\rm H})$ becomes independent of $J_{\rm H}$, and the
Schrieffer result\cite{Sch67} $J^*_{\rm K}=J_{\rm K}/2$, $T^*_{\rm K}\propto
T^2_{\rm K}(0)$, holds.\par
\begin{figure}[!htbp]
   \begin{center}
     \includegraphics[width=0.47\textwidth]{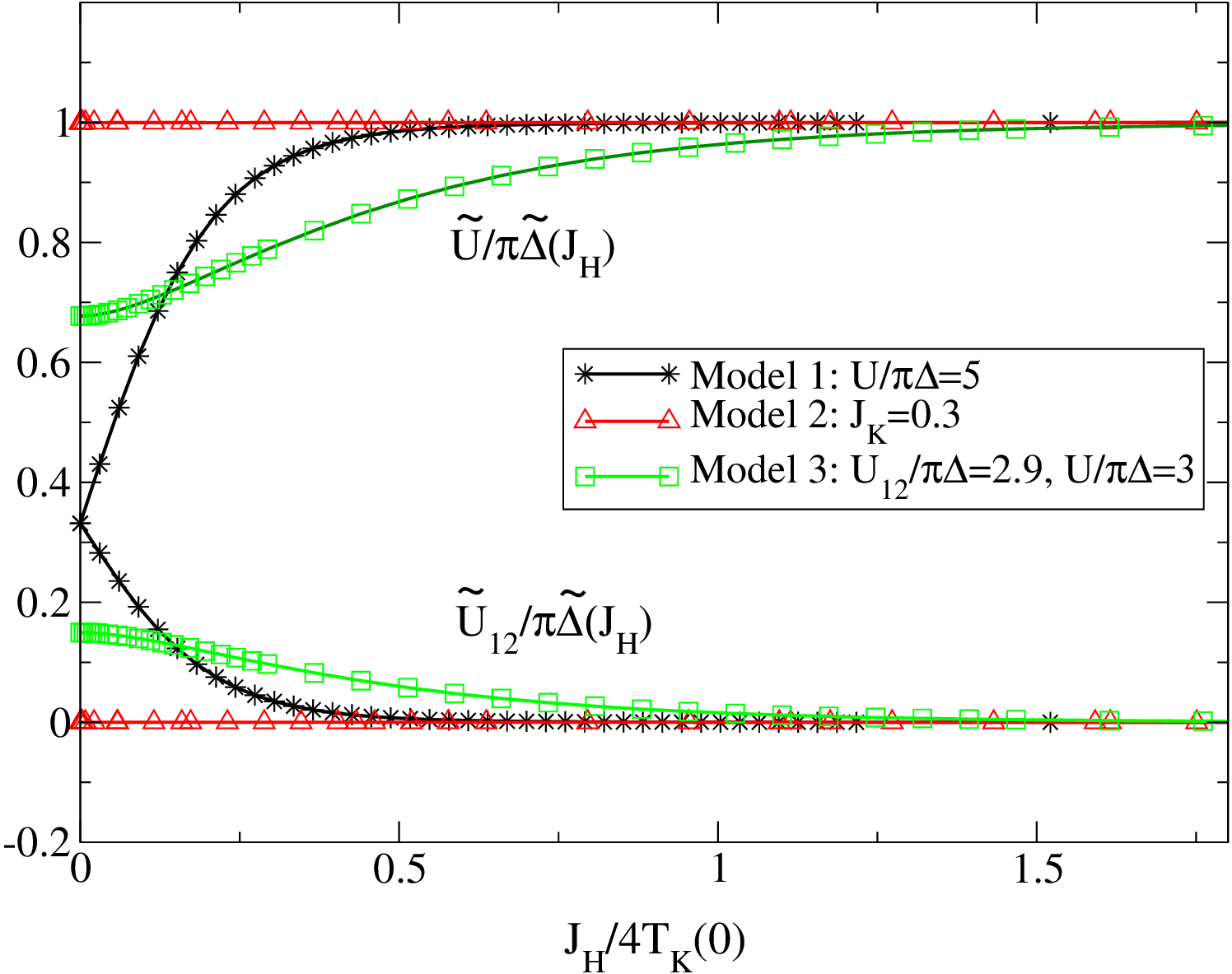}
     \caption{(Color online) A plot of $\tilde U/\pi\tilde\Delta(J_{\rm H})$
and  $\tilde U_{12}/\pi\tilde\Delta(J_{\rm H})$
versus  $J_{\rm H}/4T_{\rm K}(0)$ for the models and first three data sets shown in Fig. \ref{rdel}. The corresponding results for the last data set given in Fig.
\ref{rdel} is  $\tilde U/\pi\tilde\Delta(J_{\rm H})=1$ and  $\tilde U_{12}/\pi\tilde\Delta(J_{\rm H})=0$ for all values of $J_{\rm H}$ in the range shown.
 } 
     \label{ru_ru12}
   \end{center}
 \end{figure}

 
The remaining renormalized parameter ratios,  $\tilde U/\pi\tilde\Delta(J_{\rm
  H})$ and $\tilde U_{12}/\pi\tilde\Delta(J_{\rm H})$, that  characterize the
low energy fixed point, are shown in Fig. \ref{ru_ru12} for the different models as a function of  $J_{\rm H}/4T_{\rm K}(0)$. For model 1
with $U/\pi\Delta=5$ the charge fluctuations are suppressed $\chi_c\sim 0$, so from
Eq. (\ref{chic}) when $J_{\rm
  H}=0$ ($\tilde J_{\rm H}=0$)  for $n=2$ we get $\tilde
U/\pi\tilde\Delta=1/3$. As $U_{12}=U-3J_{\rm H}/2$ for this model when 
$J_{\rm
  H}=0$, we also predict the result  $\tilde
U_{12}/\pi\tilde\Delta=1/3$. These are confirmed in the results given in
Fig. \ref{ru_ru12}. When $J_{\rm H}$ is increased and the   orbital
fluctuations are also suppressed  we get the results for this model given in
Eq. (\ref{relation2}), which imply that in this limit $U_{12}\to 0$,
and the NRG calculations confirm these results. \par
The NRG results for model 2 give $\tilde
U/\pi\tilde\Delta=1$ and  $U_{12}= 0$ for all values of $J_{\rm H}$.
We also obtain the same results for model 3 (not shown) for the case
$U_{12}=0$, $U/\pi\Delta=5$.
This is not surprising because, as noted earlier,  when  $U_{12}\to
0$ and $U_{12}/\pi\Delta\gg 1$, model 3 and model 2 are very similar. \par
Finally in Fig. \ref{ru_ru12} results are
 shown for model 3 for the case $U_{12}/\pi\Delta=2.9$ and
$U/\pi\Delta=3$. To test the predictions in this case, the expression for 
 $\chi_c$ given in
Eq. (\ref{chic}) must be generalized as it assumes the relation.
$U_{12}=U-3J_{\rm H}/2$. As both charge and spin are conserved in model 3,
from the 
corresponding Ward indentities, exact results for the  charge and  spin
susceptibilities can be derived.  For $n=2$ the generalized expression for the
charge susceptibility takes the form, 
\begin{equation}
\chi_c=4\tilde \rho^{(0)}(0)(1-\tilde U-2\tilde U_{12}
)\tilde\rho^{(0)}(0)).
\label{chic2}
\end{equation}
As the charge fluctations are suppressed for model 3
with  $U_{12}/\pi\Delta=2.9$ and
$U/\pi\Delta=3$, Eq. (\ref{chic2}) predicts 
the result $\tilde U/\pi\tilde\Delta+2\tilde U_{12}/\pi\tilde\Delta=1$,
which is satisfied in the results shown in Fig. \ref{ru_ru12}
for all values of $J_{\rm H}$. Though both  $\tilde U/\pi\tilde\Delta$ and
$\tilde U_{12}/\pi\tilde\Delta$ are non-zero for  $J_{\rm H}=0$,
 as $J_{\rm H}$ is increased  $\tilde U/\pi\tilde\Delta\to 1$, $\tilde
U_{12}/\pi\tilde\Delta\to 0$ in line with the result in Eq. (\ref{relation2}).\par

\vspace*{0.7cm}
 \noindent
\vspace*{0.7cm}
 \noindent

\begin{figure}[!htbp]
   \begin{center}
     \includegraphics[width=0.47\textwidth]{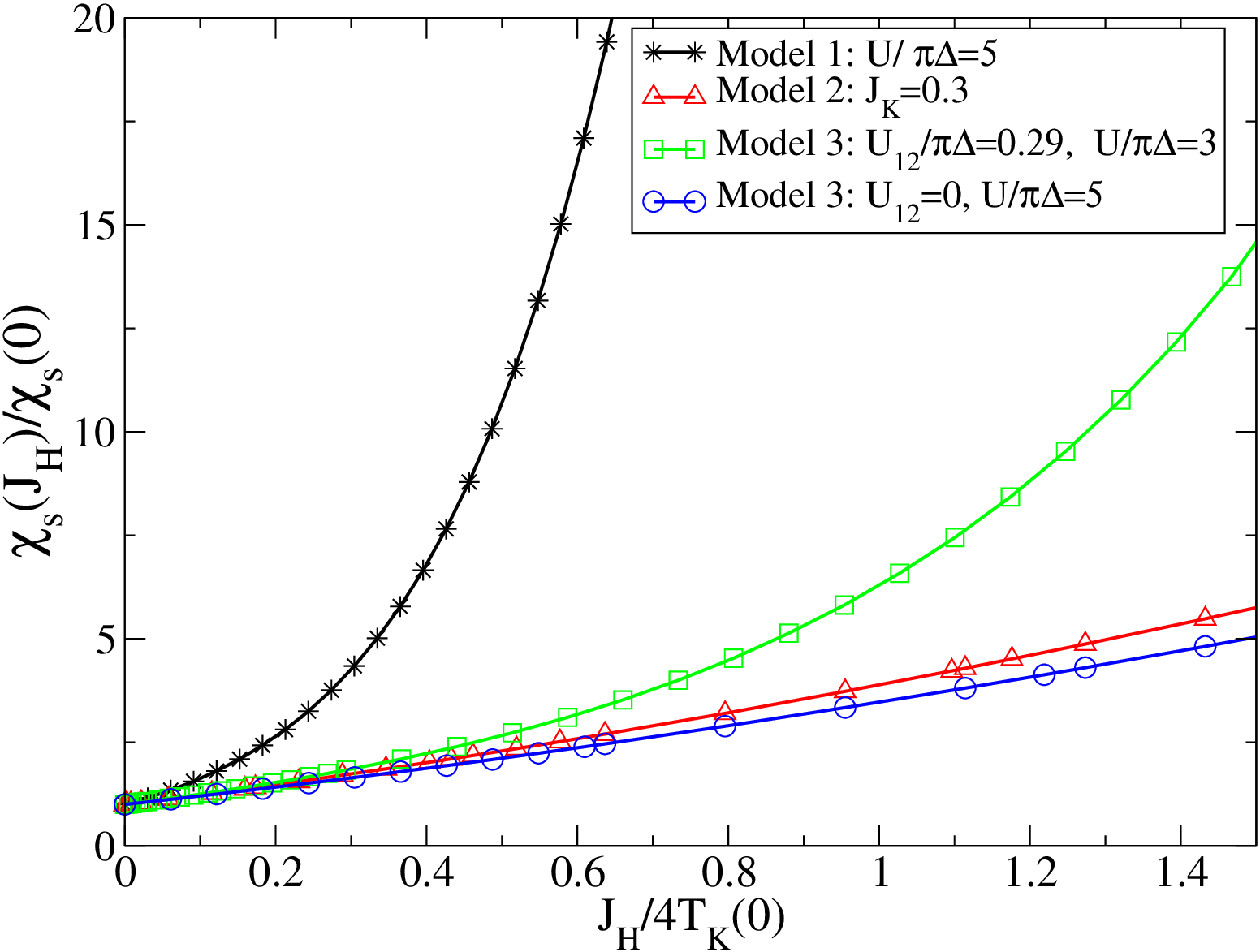}
     \caption{(Color online) A plot of $\chi_s(J_{\rm H})/\chi_s(0)$
versus
  $J_{\rm H}/4T_{\rm K}(0)$ for the models for the data sets shown in Fig. \ref{rdel}. 
 } 
     \label{chisf}
   \end{center}
 \end{figure}

The expressions for spin susceptibility $\chi_s$  and Wilson ratio $R_{\rm W}$
for model 3 with
$U_{12}\ne U-3J_{\rm H}/2$ remain  unchanged  from those given in 
 Eqs. (\ref{chis})  and (\ref{WR}). To calculate the
impurity spin susceptibility for the models and parameters sets used in the other
plots  we substitute the renormalized
 parameters into Eq. (\ref{chis})
 and give the results in Fig. \ref{chisf}. The exponential decrease of the Kondo temperature with increase of
$J_{\rm H}$ gives a dramatic rise in $\chi_s$ for model 1 compared with the
corresponding results for models 2 and 3.
However, the change in the corresponding  Wilson ratios  on increasing  $J_{\rm
  H}$ from zero is much less dramatic. For model 1 with $U/\pi\Delta=5$, 
$R_{\rm W}$ increases from $4/3$ for  $J_{\rm
  H}=0$ to  $8/3$ for large $J_{\rm
  H}$, and for model 2 from $2$ to  $8/3$ over the same range.\par

\noindent
\vspace*{0.7cm}
 \noindent

 Yoshimori \cite{Yos76} 
derived an exact result for the low temperature 
 impurity contribution to the resistivity for model 1 in the  particle-hole symmetric
case and  $H=0$. In terms of the renormalized parameters, the result is
\begin{equation}
R(T)=R_0\left(1-{\pi^4(1+I)\over 48} \left({T\over T_{\rm K}}\right)^2+{\rm O}(T^4)\right),
\label{res1}
\end{equation}
where $I$ is given by 
\begin{equation}
I= 2((2n-1)\tilde U^2
-6(n-1)\tilde J_{\rm H}(\tilde U-\tilde J_{\rm H}))/(\pi\tilde\Delta)^2,
\label{IR}
\end{equation}
and $\pi\tilde\Delta=4T_{\rm K}$. For models 2 and 3 for the case $n=2$ this
result can be generalized using renormalized perturbation to second order
in the renormalized interaction vertices for the impurity self-energy\cite{NCH10s,NCH12a,NCH12b} to give 
\begin{equation}
I=  (2\tilde U^2 +4\tilde U_{12}^2+3\tilde J^2_{\rm H})/(\pi\tilde\Delta)^2,
\label{IR2}
\end{equation}
which takes the value $I=10/3$ in the regime where Eq. (\ref{relation2}) holds.\par

\section{Conclusions}
We have shown that we can get very different results for the Kondo
temperature as a function of the Hund's rule coupling $J_{\rm H}$ depending on
how this term is included in a model of a magnetic impurity. The physically
most relevant model for a magnetic impurity with $n$-fold degenerate states in the absence of
crystal field or spin-orbit splitting should be the one introduced by
Yoshimori \cite{Yos76} which conserves orbital angular momentum about the impurity site,
or the generalizations of this model considered by Mih\'aly and
Zawadowsky\cite{MZ78}, and  Yoshimori and
Zawadowski \cite{YZ82}, where off-diagonal matrix elements
of the scattering 
between  orbital  channels are included,  subject to the condition of
overall conservation of angular momentum. These matrix elements omitted
from the Yoshimori model have been shown by  Nozi\'eres and Blandin \cite{NB80}
not to affect the results in low temperature  Fermi liquid regime,
nor the derivation of the impurity spin  and orbital susceptibilities
based on  the Ward identities deduced from  conservation of spin and 
orbital angular momentum\cite{YZ82}. The restriction to models where angular momentum
is conserved about the impurity site might not be appropriate
to all situations, such as use of the model 3 (given in Eq. (\ref{model2}))
 to describe the interaction
between two quantum dots\cite{SNOHT12}. We have studied in particular the model
used by Nevidomskyy and Coleman \cite{NC09} for the case $n=2$
 and, for the smaller values of $J_{\rm K}$, verified  their scaling equation  $T_{\rm K}\propto
1/J_{\rm H}$ when the Hund's rule coupling becomes strong enough to 
lock the individual spins to form an effective spin 1 coupled to the 
conduction electrons with an effective Kondo coupling $J^*_{\rm K}$. We also found
the point  $J_{\rm H} \sim J_{\rm K}$ where this scaling crosses over to the
regime considered by Schrieffer\cite{Sch67}, where $J^*_{\rm K}=J_{\rm K}/n$
and the  $T_{\rm K}$ becomes independent of  $J_{\rm H}$.\par
We find that the low temperature behaviour of all the models considered  can be described in terms of a
quasiparticle Hamiltonian with renormalized parameters, which,   for the case
$n=2$, can be deduced
from an analysis of the low energy fixed point of the NRG calculation.
We have shown that, despite the differences between the models, in the
strong Hund's rule regime, just two interaction parameters are required,
a renormalized interaction $\tilde U$, within each impurity orbital,
and a renormalized Hund's rule interaction between electrons in different
orbitals $\tilde J_{\rm H}$, which can both be expressed in terms of the
Kondo temperature, $\tilde U=4T_{\rm K}$ and  $\tilde J_{\rm H}=8T_{\rm
  K}/3$. 
\par

\noindent{\bf Acknowledgment}\par
\bigskip\par
{We thank Daniel Crow and Johannes Bauer for helpful discussions. This work has been supported in part by the EPSRC Mathematics Platform grant EP/1019111/1.}
\par

\bibliography{artikel}
\bibliographystyle{h-physrev3}
\end{document}